# The Dynamics of the Galactic Bar

Martin D. Weinberg[1]

*Department of Physics & Astronomy, University of Massachusetts, Amherst, MA 01003*

## 1. Introduction

The dynamics of formation and evolution of structure in barred galaxies is subtle and will require many and detailed observations to discriminate between the alternative hypotheses. Why should someone interested in such problems consider the Milky Way? In terms of data volume, our knowledge of the Milky Way is vast and the availability of detail is its major advantage. In principle, one can study morphological details such as orientations, strengths of asymmetries and kinematics details such as velocity field/pattern speeds using a wide variety of tracers. To illustrate, theorists have not converged on a single mechanism to explain bars (witness the *instability* vs. *secular formation* debate). It is possible that both operate in different regimes depending on internal or external influences: internally, triaxialities and misalignments in the bulge, spheroid or halo can apply torques and drive angular momentum waves which saturate to form a bar; and externally, satellite galaxies can exchange orbital angular momentum with its disturbance which has the similar effect. In short, if the Galaxy is indeed barred, it may hold clues to some of the detailed problems posed at this meeting.

This review has several goals. In the previous paper, Konrad Kuijken presented the evidence for a barred Milky Way. When one thinks of the Milky Way as a barred galaxy, one probably has a mental picture similar to the artist's conception by De Vaucouleurs & Pence (1978, their Fig. 6). How content should we be with this picture? To this end, I will selectively review the various dynamical scenarios that have been explored to date, especially those that illustrate the conundrums (§2.). In short, although the existence of asymmetries are convincing enough, the interpretation remains ambiguous. You will see that their explanation is woefully incomplete and hopefully the barred galaxy community will find the details intriguing enough to remedy this. In addition, I was struck by lack of coherent picture for the Milky Way asymmetries as I prepared for this meeting; each mechanism is considered independently of all others. With this motivation, I will present (§3.) a nonstandard picture which has the potential to explain many of the signatures and a few of the conundrums. Moreover, it illustrates global features are dynamically connected and I believe that it is time to revisit the Milky Way in this context. Finally, I will end with a wish list for future work—both observational and theoretical—designed to help us pin down the Milky Way.

---

## 2. Dynamical models

Because of the indirect nature of most bar detections, researchers have adopted the generic bar scenarios to infer the existence of the features. There appears to be asymmetric features on at least three scales:

1. Nuclear—inside of 1 kpc;
2. Inner—inside of 5 kpc;
3. Outer—at or outside of the solar circle.

In this section, I will review the dynamics behind some of these inferences before discussing the bigger picture in §3.. Although these models are success stories, we must guard ignoring minor disagreements which may be major clues. Space constraints rule out a comprehensive survey and I apologize in advance for covering only major themes.

### 2.1. Nuclear bar

A number of studies (e.g. Teuben et al. 1986, Mulder & Liem 1986, and Athanassoula 1989, 1992ab, see Teuben in this volume) have given us faith that the basic features of gas flow in a non-axisymmetric potential can be understood by studying closed orbits. This method works for galaxy models with mostly regular orbits for which the streamlines remain close to stable families of particle orbits. Where families exchange stability, orbits cross resulting in rapid evolution until a new smooth flow is established. With this principle in hand, Binney et al. (1991) found remarkable agreement between the HI $l$–$v$ diagram inferred from closed orbits in a rotating potential about the galactic center. The identification relies on the location of self-intersecting $x_1$ orbits[1] and the model therefore depends critically on pattern speed. Binney et al. (1991) predict a corotation radius of 2.4 kpc.

However, if you look closely at the HI contours (see Blitz et al. 1993), you will notice that there is more emission in the first quadrant. In fact, Binney (1993) has pointed out that this discrepancy is even more pronounced in CS and that Bally et al. (1988) report that 3/4 of the molecular gas within $|l| < 2°$ is in the first quadrant. Similar offsets have been noted in the position of the putative Galactic Center itself, Sgr A*. Blitz (1995) reported that a reanalysis of the HI, CO and COBE data imply a centroid at $l \approx 0.5°$.

Theoretically, should we expect the centers of galaxies to be centered? Weinberg (1994b) argued that the centers of hot stellar systems—such as bulges, spheroids, star clusters—are easily displaced with a very weakly damped *sloshing* or *seiche* mode. This causes an offset central density peak which rotates slowly (compared to an orbital period) about the initial kinematic center. Recent work (paper in preparation) shows the same dynamics applies to disks (see Fig. 1). I conclude that we *should* expect offsets, although an exhaustive exploration of their amplitude for realistic scenarios is required. A possible mechanism will be discussed in §3..

---

[1] see Contopoulous & Papayannopoulos 1980 for details

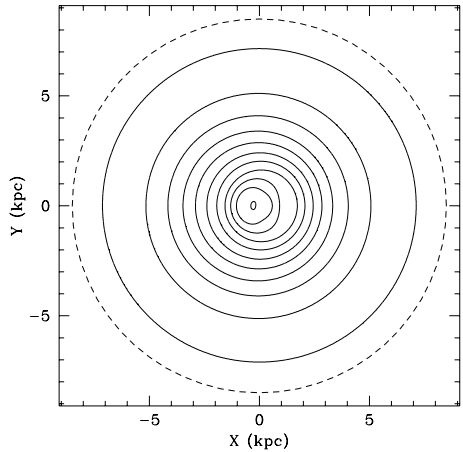

Figure 1. Density contours for an exponential disk with a weakly damped $m = 1$ mode embedded in an Hernquist model halo/spheroid of equal mass and scale lengths. Note the offset at hundreds of parsecs and nearly circular contours beyond several kiloparsecs.

## 2.2. Classic stellar bar

Detecting our own stellar bar is difficult and indirect. The previous reviewer addressed the evidence for both a bulge bar (e.g. Blitz & Spergel 1991b, Weiland et al. 1994, Dwek et al. 1995) and disk bar. The main qualitative difference between these suggestions and those in §2.1. is that these inferences are photometry-based. Generally, such morphological identifications have dynamical implications which lead to distinct kinematic predictions.

I will take an example from my own work which applies equally well to other cases. Based on IRAS-selected AGB stars, Weinberg (1992) inferred bar with half-length 2–4 kpc. A recent analysis which rigorously treats selection effects and censored data confirms this result (Nikolaev & Weinberg 1995). In addition, it estimates a 3.5 kpc scale length for the AGB star candidates, strengthening their identification with the old disk population. As discussed earlier in this meeting, bars end near or inside corotation; let us take this inferred length to be 3 kpc.

Now as we know, the pattern speed and gravitational potential determines the location, morphology and stability of resonant orbit families. Therefore, a given bar model leads to distinct kinematic predictions and provides a way to check each scenario. Details were presented in Weinberg (1994a) who found that both the correlated deviation of the stellar velocity field from circular and increased velocity dispersion near the resonances are potentially observable (his Figs. 12& 14). In addition, if the bar is secularly evolving by interaction with disk or spheroid, the kinematic signature is modified. Under most circumstances, the spheroid will remove angular momentum (Weinberg 1985, Hernquist & Weinberg 1992) and cause the bar to slow (Little & Carlberg 1991). This causes orbit trapping near resonance which can be detected. Figure 2 shows the guiding centers of orbits trapped about OLR and their line-of-sight signature for a bar with constant (left) and decreasing (right) pattern speed. Orbits appear in orthogonal orientations for the same guiding center radii for a changing pattern speed.

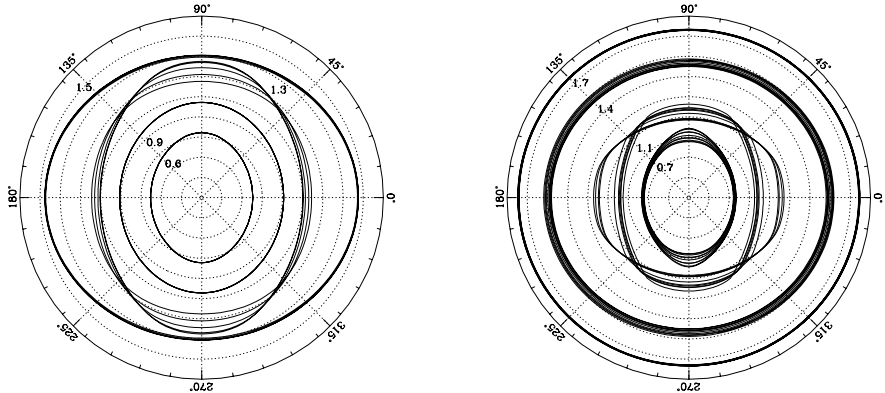

Figure 2. Mean trajectories for ensembles of orbits initial guiding center radii as labeled. Ten different initial phases are shown for each guiding center. Bars with fixed pattern speed ($R_{ILR} = 1.3$, left) and decreasing speed ($R_{ILR} = 1.1 \rightarrow 1.3$, right). Phase independence is broken by trapped orbits causing some ensembles to spread.

The same arguments apply to a rotating bulge bar. In fact, it is possible that both disk and bulge bars are part of the same dynamical system which is supported by the apparent alignment of the major axes. If so, the rough alignment of the bulge with the nuclear bar is fortuitous. In favor of this interpretation, the 3 kpc expanding arm has been nicely modeled by Yuan (1993) and requires even a faster pattern speed than inferred by Binney et al.: corotation at 1.6 (1.2) kpc for a two- (one-) armed spiral. Another possibility for the 3 kpc arm that has not been explored is that the Milky Way's nuclear bar is offset (cf. §2.1.) causing a one-armed gas disturbance as described by Colin & Athanassoula (1989). Clearly, additional kinematic study is badly needed and should help resolve these uncertainties.

## 2.3. Outer bar/triaxial spheroid

Based on asymmetries in the global HI $l$–$v$ diagram, Blitz & Spergel (1991a, BS) postulated a slowly rotating spheroid (ILR outside of the solar circle) which gives rise to an outward LSR motion of approximately 14 km s$^{-1}$. This interpretation followed their finding that gas motion in outer Galaxy is consistent with circular and their failure to find a closed orbit solution for stationary LSR which reproduced the observed asymmetries.

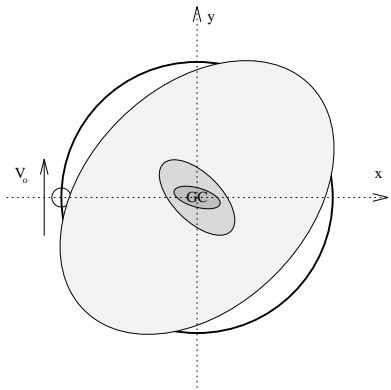

Figure 3. The scale and orientation of reported Milky Way asymmetries listed from inside outward: bulge or HI-inferred bar, disk bar, triaxial spheroid.

More generally, this work illustrates the strong constraint imposed by well-determined local kinematics. These issues have been addressed in the context of geometric asymmetries by many and recently summarized by Kuijken & Tremaine (1994, KT). Based on a literature survey, they report a radial LSR velocity of $-1 \pm 9 \,\mathrm{km\,s^{-1}}$. On the other hand, stellar velocities toward the anticenter based on distant stars suggest recession at $\sim 10 \,\mathrm{km\,s^{-1}}$. As pointed out by Kuijken (1992), we are left with the conundrum that the stars and gas seem to be moving in opposite directions. In addition, a non-zero vertex deviation is potentially and indicator for non-circular asymmetries. The BS model predicts $l_v = -9.3°$ whereas KT report $l_v \approx 5.5 \pm 4.2°$. (See KT for other quantities). This tentatively suggests rejection of the BS model in its current form, but does not resolve the mystery. Overall, one concludes that local kinematics do not give strong hints of asymmetries, but conversely, are very useful tests for any global interpretation.

At this point, we are left with an inferred picture of the Milky Way shown in Figure 3. There appear to be 2 maybe 3 alignments or anti-alignments. Is this coincidental or is there a conspiracy at work?

### 2.4. Halo

My speculative answer in §3. implicates the unseen component and I briefly list halo properties below to set the stage. Since the halo is the most massive of Galactic components and is capable of supporting and possibly sustaining structure, it almost certainly plays an important role in the dynamics of the entire Galaxy.

Mass determinations have a long history. Recently, Little & Tremaine (1987) developed a Bayesian mass estimator which yielded the remarkable values $M_h \approx 2.5 \times 10^{11} \,\mathrm{M_\odot}$ and $R_{edge} \lesssim 50\,\mathrm{kpc}$ based on the positions and velocities of distant satellites. Soon after, Zaritsky et al. (1989) applied the same techniques using new measurements for the Leo dwarfs and found $M_h \approx 12.5 \times 10^{11} \,\mathrm{M_\odot}$ us-ing Little & Tremaine's method and timing arguments. Norris & Hawkins (1991) estimated $M_h \sim 10^{11}$ inside of $R \sim 40$ kpc with no sign of turn over in the rotation curve using halo star counts. Lin et al. (1995) predict $M_h \sim 5.5 \pm 1 \times 10^{11}$ M$_\odot$ within 100 kpc based on theoretical models for the LMC orbit. There is general convergence on a massive halo producing a flat rotation curve at least out the Magellanic clouds and most likely a factor of two farther. Finally, we can compare the Milky Way to normal galaxies by using satellites as mass estimators. Zaritsky et al. (1993) report typical (expectation) values of $M_h = 2 \times 10^{12}$ M$_\odot$ inside of $R = 200$ pc.

Now, what about the satellite companions themselves? Early work by Holmberg (1969) suggested that around spirals satellites are rare. Based on their selected sample, Zaritsky et al. (1993) find that late-type spirals similar in luminosity to the Milky Way have 1.4 companions on average and derive a satellite luminosity function. The LMC is well within the normal range in both luminosity, although on the large side, and separation from its primary.

It appears that the Milky Way is a typical galaxy: it has a massive halo and a substantial satellite. A full dynamical picture of the Milky Way ought to include fully these two *massive* components of the Milky Way.

## 3. Using the LMC to produce Milky Way asymmetries

Researchers have tried to use the LMC[2] to produce the warp in the HI disk in the outer Galaxy, but back of the envelope calculation tells us that the tidal amplitude will be too small to produced the observed effect (e.g. see Binney 1992). The new twist here is to use the LMC to perturb the halo and let the halo distort the disk. This has implications for the warp too, but I will not discuss them here.

The location and orientation of the LMC orbit is based on the recent proper motion analysis by Jones et al. (1994). Using their space velocity, the location of the clouds, and assuming a spherical isothermal halo with IAU values for $R_o$ and $V_o$, one can infer the orbital plane. Taking the galactocentric coordinate system with $\hat{x}$ along the Sun–Center line, $\hat{y}$ in the direction of LSR motion and $\hat{z}$ toward the NGP, the LMC orbital plane has latitude $76 \pm 13°$, longitude of ascending node $-82 \pm 10°$, and argument of the perigalacticon $-36 \pm 3°$[3]. More detailed orbit models based on this proper motion and the position of the stream suggest that $M_h \approx 5 \times 10^{11}$ inside of 100 kpc (Lin et al. 1995) and this will be adopted here. Inside of this halo, I embedded an exponential disk with scale length $a = 3.5$ kpc and $M_{disk} = 6 \times 10^{10}$ M$_\odot$ based on $L = 1.2 \times 10^{10}$ L$_\odot$ and $\Upsilon = 5$ (Binney and Tremaine 1987). The disk self-consistently feels the gravitational field from its own and the halo's mass distribution. The distribution function for the disk is constructed using a quadratic programming method, similar to that proposed by Dejonghe (1989). This gives a flat rotation curve comparable to the observed estimates. For the mass of the LMC, I take the conservative estimate based on Meatheringham et al. (1988) of $M_{LMC} = 6 \times 10^9 \, M_\odot$. A value of $M_{LMC} = 1.5 \times 10^{10} \, M_\odot$, nearly a factor of three larger, by Schommer et al. (1992) is based on an HI rotation curve.

---

[2]For the purposes of this talk, I will lump the Magellanic Clouds together and call them both the LMC.

[3]The error bars are based on Gaussian error propagation and are dominated by those of the Jones et al. space velocity.

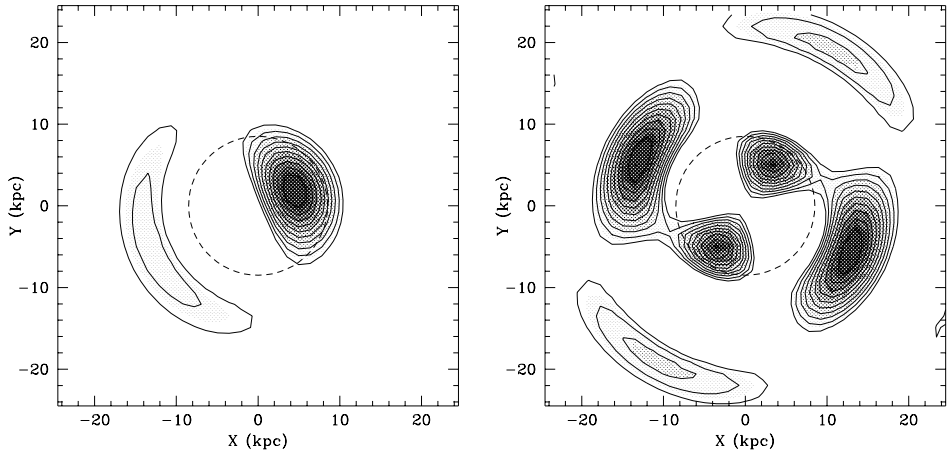

Figure 4. Contours of halo overdensity in the orbital plane for an eccentric orbit ($\epsilon = 0.4$) with pericenter at $X = -50$ kpc, $Y = 0$ kpc and $\mathbf{V} = -\hat{Y}$ direction. The $m = 1$ (left) and $m = 2$ (right) components are shown along with location of the solar circle (dashed). The peak $m = 2$ response is 1% of the peak $m = 1$ response.

The LMC orbit is decaying due to dynamical friction (e.g. Lin et al. 1995) and this produces a wake (e.g. Weinberg 1989). The $m = 1$ and $m = 2$ wake components[4] demonstrate the non-local interaction (Fig. 4): the satellite at $(X, Y) = (-50 \, \text{kpc}, 0 \, \text{kpc})$ does produce a local wake but the strongest distortion is inside of 20 kpc. Although most of the *mass* in the wake in the outer Galaxy, the peak *density* response is near the solar circle.

We can now determine the response of the disk to the LMC. The halo and disk respond to the LMC, their own wakes, and each other's wakes. Figure 5 depicts the distortion of the disk by the halo and the LMC; both absolute and relative density scales are shown. The dominant response is near the solar circle and has significant amplitude: several percent at the solar circle reaching 20% at $R_g \approx 12.5$ kpc and 30% near $R_g \approx 3.5$ kpc. This should lead to a detectable offset between the inner and outer galaxy. Finally, note that the LMC has a retrograde orbit relative to disk rotation, and the response has a negative pattern speed relative to the disk rotation. Therefore, the slow retrograde patterns in Figure 4 can not be directly responsible for classic prograde disk features but could be the triggering source.

---

[4]Because the polar harmonics $m$ couple to all spherical harmonics $l \geq |m|$, the response in Fig. 4 has been summed over $l \leq 6$.

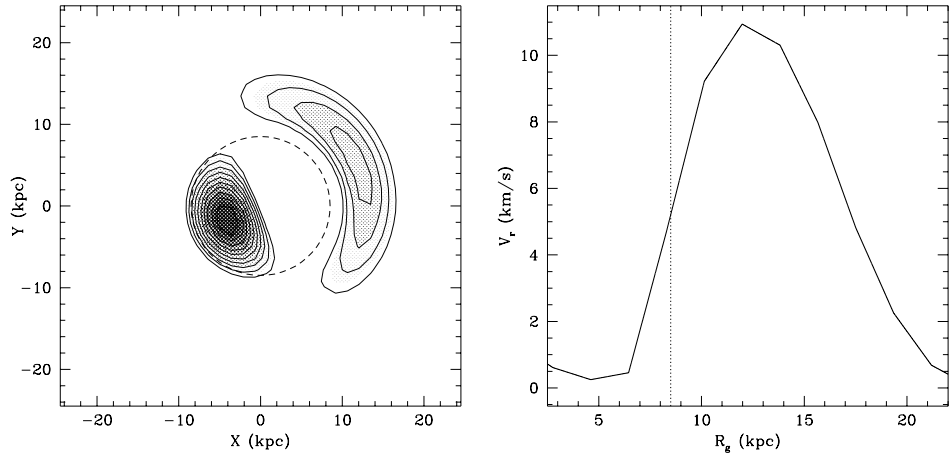

Figure 5. Positive overdensity for the $m = 1$ disk response (shaded). Contours indicate relative overdensity spaced linearly from 3% to 30%. The Sun has coordinates $(-8.5\,\text{kpc}, 0\,\text{kpc})$.

Figure 6. Line-of-sight velocity for the $m = 1$ response viewed from the galactic center. Values are scaled to the Milky Way: scale length $a = 3.5\,\text{kpc}$ and rotation speed $V_o = 220\,\text{km s}^{-1}$.

Since the solution to the Boltzmann equation yields a distribution function, it is straightforward to compute the velocity perturbations. Figure 6 the line of sight velocity from the center in the direction of the anticenter. Interestingly, the relative velocity between the LSR and the outer galaxy is roughly $6\,\text{km s}^{-1}$, in fair agreement with the observation of K giants and carbon star towards the anticenter. Although not illustrated here, a similar calculation shows that the amplitude of vertex deviation produced by this model is $\lesssim 1°$.

To summarize, the LMC perturbation is consistent with the local kinematics, produces the observed trend in anticenter kinematics, and at least hints at some of the structure discussed in §2.. A number of details remain to be worked out. First, the pattern speed of this disturbance is very slow and retrograde, $-4.3\,\text{km/s/kpc}$, and therefore can not directly be the source of an inner bar. However, the LMC disturbance could be the source of disk noise which is then swing amplified or could excite the observed central offset. Second, the asymmetries in the $l$–$v$ diagram that Blitz & Spergel (1991a) sought to explain may fit into to this picture. Because the wake is time-dependent, one might expect a difference between stellar and gas response. A simple closed-orbit analysis will not be sufficient and an explicit check of the hydrodynamic response in the presence of the halo disturbance will be necessary. Finally, Figure 5 suggests that the LMC wake near pericenter at least, can produce a global disturbance of several tens of percent inside the solar circle; is this enough to trigger bar formation? Once formed, the bar may be independent of the original pattern, similar to demonstrations by Combes (1993) and Friedle & Martinet (1993) that that a nuclear bar can decouple from the embedding disturbance. And, of course, the same mechanism may be able to preserve the convincing signature of Binney et al. even in the larger-scale bar pattern.

## 4. Wish list

This brings me to my list of topics which cry for attention:

- **Observe pattern speeds.** The scenarios described in §2. and §3. each predict or at least imply different pattern speeds. Each pattern speed is likely to have a distinguishing kinematic signature. Clearly, we need more tracer observations, especially at large scales, in order to discriminate between hypotheses. I advocate AGB stars because they represent the old disk and are intrinsically bright infrared targets where interstellar extinction is weaker. They may be classified spectroscopically and selected by variability. 2MASS should detect $\gtrsim 10^5$ candidates and a pilot spectroscopic classification and radial velocity program is in progress. Cepheids share the same advantages but represent the young disk population. Finally, K-giants studies, the workhorse of galactic structure, will continue to be useful. (Also see Konrad Kuijken's review in this volume.)

- **Understand the dynamics of stars** *and* **gas.** The dynamics of stars and gas are different and different responses to a disturbance might be expected. Although common, periodic orbit analysis may be a poor indicator of gas streamlines in the outer galaxy both because the galaxy may be too young to have phase mixed its transients and because the outer galaxy is subject to both perturbations from its satellites as discussed in §3. and possibly other interlopers. Conversely, simultaneous analysis of stars and gas can be exploited to discriminate between theories.

- **Study interactions between components.** Don't forget about the halo! We have two examples now of interactions with the halo that are likely to drive or modify evolution: the bar–halo interaction and the satellite halo interaction. Because gravitationally coupled disturbances are global, components can cause their mutual evolution and external disturbances can be transmitted at all scales. Given that halos can sustain structure, it would not surprise me if, converse to the Ostriker-Peebles criterion, we find that can halos *CAUSE* bars.

## 5. Summary and conclusions

I hope to have raised the possibility, if not convinced you, that structure in the halo/spheroid and disk are inseparable. In the outer galaxy, perturbations by satellites and encounters with neighbors excite structure in the halo which affects the disk. As an example, we explored the LMC as a mechanism for observed asymmetries and local kinematics and future work will extend this to HI warp.

Features with well-defined pattern speeds, most notably bars, will create distinct kinematic features. Photometry and kinematic data together is a statistically stronger diagnostic for predicted dynamical scenarios than either alone. Similarly, because gas behaves differently than stars and can produce structure at different scales, the hypothesis test is made even stronger by adding gas information.

Finally, the study of Milky Way structure and general galactic structure is complementary. For example, the morphological type of a regular face-on spiral is a directly observable quantity, but for the Milky Way, morphological type relies on an elaborate chain of inference. Conversely, a detailed test that is best posed as a star count statistic, such as the radial velocity dispersion near a resonance, may be all but impossible for any galaxy but the Milky Way. We stand to learn a great deal by analyzing both together!

**Acknowledgments.** I thank the IAU for travel support. This work was supported in part by NASA grant NAG 5-2873 and the Sloan Foundation.

**Discussion**

*Daniel Pfenniger*: Don't you think that the *non-linear* response may lead rapidly to much different pattern speeds in the central regions of your halo-disk perturbed model?

*M. Weinberg*: Yes, I expect the halo will only trigger disk structure, especially in the inner disk, rather than force it at some uncharacteristically slow pattern speed. Although subsequently, the halo may modify the evolution of disk structure.

*R. H. Miller*: Can you perturb the halo by the LMC (over the life of the Galaxy), let the disk respond, but still keep the velocity dispersions of the disk stars as low as they are in the solar neighborhood?

*M. Weinberg*: Because my disk component is constrained to two dimensions, I can not give you an answer based on calculation. However, my guess is that the slowly varying, large-scale distortions under discussion will not produce the significant disk heating that they do in disk-dwarf mergers. But, I do plan to look at this in the future.

*K. Z. Stenek*: You mentioned the possible offset between the center of the Galaxy and the center of the bar/nuclear bar. Maybe we don't know the center of the Milky Way precisely enough?

*M. Weinberg*: I agree that the notion of a center is ambiguous. However, I believe that if one defines the kinematic center through the HI rotation curve, the roughly $0.5°$ offset is well beyond the uncertainties (see Blitz 1995 for more detail).

*K. Freeman*: On the LSR streaming and identification of the OLR: Wilson (unpublished) traced the two local star streams out of the solar neighborhood, for approximately 500 pc. He showed that the star streams can be followed nicely but *only* for the younger ones with [Fe/H] in a tight range (+0.2 to -0.1 approximately). So I don't think one should over interpret this identification with the OLR.

*M. Weinberg*: That is a fair objection, both for an OLR from an inner bar or for an ILR from a triaxial spheroid. Along similar lines, Paul Schechter has claimed that the apparent LSR motion deduced from HI might be more naturally understood as a nearby spiral arm. I certainly agree that the proper approach is to allow the observations to rule out hypotheses that over produce LSR motion rather than select a "best fit" model.

*Tony Garcia-Barreto*: Would you predict an offset between the dynamical and kinematic centers of external galaxies?

*M. Weinberg*: Yes, but at this point it is a speculation based on the appearance of the mode (cf. Fig. 1) in a variety of models with small damping rates. Before predicting the expected magnitude and frequency of offsets, I need to compute their amplitude under a variety of realistic scenarios.

*B. Elmegreen*: Do the models of the LSR first remove the expected streaming motions in both gas and old stars from the known spiral arms?

*M. Weinberg*: I do not know of any model which attempts this. The issue underscoring your question (and Ken Freeman's comment) is the uncertainty in the inferred LSR motion caused by a relatively local disturbance like spiral arm or association. It is worth bearing in mind.

*J. Palous*: In my opinion, the identification of stellar streams (Kapteyn's) with orbits near OLR may be an over interpretation (of OLR is near the Sun). You would run into problems with the lack of evidence in the velocity ellipsoid.

*M. Weinberg*: I accept that the interpretation risky because the signature may not be unique given the current data. However, it will not conflict with the local velocity ellipsoid provided that the resonance is far enough away that trapped orbits do not intersect the solar circle. In that case, the mean guiding center trajectories are non-circular but the velocity dispersions are relatively unaffected.